\documentclass{llncs}

\usepackage{epsfig}
\usepackage{listings}
\usepackage{float}
\usepackage{placeins}
\usepackage{balance}
\usepackage[utf8x]{inputenc}
\usepackage[T1]{fontenc}
\usepackage{subfigure}
\usepackage{tabularx}
\usepackage{color}
\usepackage{colortbl}
\usepackage{multirow}
\usepackage[plainpages=false,pdfpagelabels=true,pdftitle={The Abandoned Side of the Internet: Hijacking Internet Resources When Domain Names Expire},pdfauthor={Johann Schlamp, Josef Gustafsson, Matthias Waehlisch, Thomas C. Schmidt, Georg Carle},bookmarks=true,bookmarksopen=true,pdfpagemode={UseOutlines},pdfstartview={FitV},pdfpagelayout={SinglePage},pdfborder={0 0 0},filecolor=black,linkcolor=black,urlcolor=black]{hyperref}
\usepackage{breakurl}
\usepackage{color}
\usepackage{moresize}
\usepackage{enumitem}
\usepackage{xspace}
\usepackage{booktabs}

\let\orgautoref\autoref
\renewcommand{\autoref}
{\def\sectionautorefname{Section}%
\orgautoref}

\begin{document}

\title{The Abandoned Side of the Internet:\\Hijacking Internet Resources When Domain Names Expire}
\author{Johann Schlamp\inst{1}, Josef Gustafsson\inst{1}, Matthias Wählisch\inst{2},\\Thomas C. Schmidt\inst{3}, Georg Carle\inst{1}}
\institute{Technische Universit\"at M\"unchen\\
\email{\{schlamp,gustafss,carle\}@net.in.tum.de}
\vskip 0.1cm
\and
Freie Universität Berlin\\
\email{\{m.waehlisch\}@fu-berlin.de}
\vskip 0.1cm
\and 
HAW Hamburg\\
\email{\{schmidt\}@informatik.haw-hamburg.de}
}
\maketitle
\pagestyle{empty}

\begin{abstract}

The vulnerability of the Internet has been demonstrated by prominent IP prefix
hijacking events. Major outages such as the China Telecom incident in 2010
stimulate speculations about malicious intentions behind such anomalies.
Surprisingly, almost all discussions in the current literature assume that
hijacking incidents are enabled by the lack of security mechanisms in the
inter-domain routing protocol BGP. 

In this paper, we discuss an attacker model that accounts for the hijacking of
network ownership information stored in Regional Internet Registry (RIR)
databases.  We show that such threats emerge from abandoned Internet resources
(e.g., IP address blocks, AS numbers).  When DNS names expire, attackers gain
the opportunity to take resource ownership by re-registering domain names that
are referenced by corresponding RIR database objects.  We argue that this kind
of attack is more attractive than conventional hijacking, since the attacker
can act in full anonymity on behalf of a victim. Despite corresponding
incidents have been observed in the past, current detection techniques are not
qualified to deal with these attacks. We show that they are feasible with very
little effort, and analyze the risk potential of abandoned Internet resources
for the European service region: our findings reveal that currently 73
\texttt{/24} IP prefixes and 7 ASes are vulnerable to be stealthily abused. We
discuss countermeasures and outline research directions towards preventive
solutions.

\end{abstract}

\section{Introduction} \label{sec:introduction}

Internet resources today are assigned by five Regional Internet Registrars
(RIRs).  These non-profit organisations are responsible for resources such as
blocks of IP addresses or numbers for autonomous systems (ASes). Information
about the status of such resources is maintained in publicly accessible RIR
databases, which are frequently used by upstream providers to verify ownership
for customer networks. In general, networks are vulnerable to be hijacked by
attackers due to the inherent lack of security mechanisms in the inter-domain
routing protocol BGP. Real attacks have been observed in the past that led to
the development of a variety of detection techniques and eventually of security
extensions to BGP \cite{Kent2000,RFC-6480}. Common to these attacks is a
malicious claim of IP resources at the routing layer. However, claims of
network ownership can be also made at RIR level, a fact that has received
little attention so far.

In a history of more than three decades, a vast number of Internet resources
have been handed out to numerous users under varying assignment policies. Some
ASes or prefixes have never been actively used in the inter-domain routing,
others changed or lost their original purpose when companies merged or
vanished. It is not surprising that some Internet resources became abandoned,
i.e. resource holders ceased to use and maintain their resources.

In this paper, we focus on threats that emerge from abandoned Internet
resources. Currently, there is no mechanism that provides resource ownership
validation of registered stakeholders. Instead, the control over email
addresses that are stored with RIR database objects is often considered a proof
of ownership for the corresponding resources.  Our contribution is a generalized
attacker model that takes into account these shortcomings. We thoroughly
evaluate the risk potential introduced by this attack by drawing on several
data sources, and show that the threat is real. Since this kind of attack
enables an attacker to fully hide his identity, it makes hijacking more
attractive, and significantly harder to disclose. Consequently, we show that
state-of-the-art detection techniques based on network measurements are
ill-suited to deal with such attacks. Even so, these attacks have been
evidenced in practice, and should thus be taken into account by future
research.

We continue the discussion by establishing our attacker model in
\autoref{sec:model}. In \autoref{sec:resources}, we estimate the risk potential
of abandoned resources, and show that there is a real threat. As a result, we
outline an approach to mitigate this threat, and discuss limitations of related
work in \autoref{sec:research}. In particular, we outline the need for a system
that provides resource ownership validation. We conclude our discussion in
\autoref{sec:conclusion}.

\section{Attacker Model} \label{sec:model}

Conventional attacks on BGP are based on its lack of origin validation, which
allows an attacker to originate arbitrary prefixes or specific subnets from his
own AS.  We propose a new attacker model that accounts  for attackers to take
ownership of abandoned resources. In such a scenario, an attacker is able to
act on behalf of his victim, in particular to arrange upstream connectivity.
Misled upstream providers unknowingly connect one or several ASes including
prefixes of the victims as instructed by an attacker who successfully hides his
true identity. Following this model, the anonymous attacker can participate in
the cooperative Internet exchange at arbitrary places without any formal
incorrectness. In the following, we generalize a real incident to derive
preconditions that enable this kind of attack.

\subsection{Background: The LinkTel Incident} In previous work \cite{ashijack},
a corresponding attack has been observed in practice, which is known as the
\textit{LinkTel incident}. The authors studied this attack and showed that a
victim's prefixes originated from his own AS, while the victim itself abandoned
his business.  The authors reconstructed the attacker's course of action to
claim ownership of the abandoned resources. The LinkTel incident thereby
revealed a major flaw in the Internet eco-system: validation of resource
ownership is most often based on manual inspection of RIR databases. In this
context, it was shown that the attacker was able to gain control over the
victim's DNS domain, and thus over corresponding email addresses. The involved
upstream provider presumably validated that the attacker's email address was
referenced by the hijacked resources' RIR database objects. Given this proof of
ownership, the upstream provider was convinced by the attacker's claim to be
the legitimate holder of the resources. Surprisingly, the attacker captured the
victim's DNS domain by simply re-registering it after expiration.

For several months, the attacker's abuse of the hijacked resources remained
unnoticed.  By combining several data sources, 
the authors showed that the hijacked networks were utilized to send spam, to
host web sites that advertised disputable products, and to engage in IRC
communication.  After the victim recovered his business, he learned that his
networks were listed on spamming blacklists. However, the attacker's upstream
provider refused to take action at first, since the victim was unable to refute
the attacker's ownership claims.

\subsection{Preconditions for an Attack} Based on the insights gained from the
LinkTel incident, we show that the attacker's approach can be generalized.  To
enable hijacking of Internet resources, the following preconditions have to be
met: (a) Internet resources are evidentially abandoned and (b) the original
resource holder can be impersonated.

If an organisation goes out of business in an unsorted manner, these conditions
are eventually met. As a first consequence, the organisation ceases to use and
maintain its resources. If this situation lasts over a longer period of time,
the organisation's domain name(s) expire. Since day-to-day business lies idle,
re-registration and thus impersonation becomes practicable for an attacker. At
that moment, upstream connectivity can be arranged on behalf of the victim,
since face-to-face communication is not required in general. Routers can be
sent via postal service, or even be rented on a virtualized basis. Details on
BGP and network configuration are usually exchanged via email, IRC, or cellular
phone, and payment can be arranged anonymously by bank deposits or other
suitable payment instruments. Without revealing any evidence about his real
identity, the attacker is able to stealthily hijack and deploy the abandoned
resources.

\subsection{Implications} The implications of this attacker model are manifold.
First, an attacker may act on behalf of a victim, thereby effectively hiding
his own identity and impeding disclosure. This makes hijacking more attractive
as it enables riskless network abuse. It hinders criminal prosecution, and
could be used to deliberately create tensions between organisations or even
countries. Due to the lack of a system for resource ownership validation, these
attacks only depend on idle organisations or missing care by legal successors
of terminated businesses. Even after the discovery of such an attack, it is
difficult for the victim to mitigate since reclaiming ownership is the word of
one person against another at first. The LinkTel incident \cite{ashijack}
proves that this is not only a realistic scenario: such attacks are actually
carried out in practice.

The benefit of attacks based on abandoned resources can even be higher than in
the case of conventional attacks. Hijacking productive networks rarely lasts
for more than a few hours, since the victim can receive great support in
mitigating the attack.  Moreover, for most cases, the benefit is reduced to
blackholing a victim's network -- with the Youtube-Pakistan incident being a
prominent example. In addition, monitoring systems for network operators exist
that raise alarms for unexpected announcements of their prefixes. However, due
to the very nature of abandoned resources, virtually no one is going to take
notice of an attack. Our attacker model  thus accounts for stealthily operating
attackers who aim at persistently maintaining malicious services.

\section{Abandoned Internet Resources} \label{sec:resources}
We identify readily hijackable Internet resources by searching RIR databases for unmaintained resource objects. Subsequently, we distinguish between resources that are still in use, with potential for network disruption, and resources that are fully abandoned and ready to be abused stealthily. Such resources are especially attractive for attackers for two reasons. First, the resource is assigned to an organisation for operational use and thus represents a valid resource in the Internet routing system. Second, an attacker can easily claim ownership by taking control of the contact address referenced by corresponding RIR database objects, i\.e\. by re-registering a domain name. 

Consequently, we look for RIR database objects that reference email addresses with \textit{expired DNS names}. 
Since the inference of invalid domain names can also be the result of poorly maintained resource objects or typing errors, it is important to take into account recent database activities for individual resource owners, and to correlate this information with BGP activity.

The following analysis is based on archived RIPE database snapshots over 2.5 years (23~February, 2012 till 9~July, 2014). Our results are representative for the European service region only, but similar analyses can be done with little effort for other service regions, too.

\subsection{Resource Candidates from RIR Database}
RIPE, like all other RIRs, provides publicly available database snapshots on a daily basis. Most of the personally related information is removed due to privacy concerns. Some attributes, however, remain unanonymized, which we utilize to extract DNS names.

\subsubsection{Available Data Objects}
The RIPE database holds more than 5.2 million objects. These objects can be updated from the Web or via email. Most of these objects optionally hold an email address in the \texttt{notify} field, to which corresponding update notifications are sent. Despite anonymization, we found that these \texttt{notify} fields are preserved in the publicly available database snapshots, which is also the case for \texttt{abuse-mailbox} attributes. To extract DNS names, we parse these email addresses where applicable.

\autoref{table:ripe_obj} shows the distribution of stored objects by type along with the number of DNS names we were able to extract. Although we found more than 1.5 million references to DNS names, the total number of \emph{distinct} names is only 21,061. This implies that, on average, more than 72 objects reference the same DNS name. The overall fraction of objects that reference a domain name is 29.24\%, which is surprisingly high since the database snapshots are considered to be anonymized.

\begin{table}[t!]
 \centering
 \begin{tabularx}{0.75\textwidth}{Xrrr}
 \toprule
 \textbf{Object type} & \textbf{Frequency} & \multicolumn{2}{c}{\textbf{DNS references}} \\
 \midrule
 \texttt{inetnum} & 3,876,883 & 1,350,537 & (34.84\%) \\
 \texttt{domain} & 658,689 & 97,557 & (14.81\%) \\
 \texttt{route} & 237,370 & 50,300 & (21.19\%) \\
 \texttt{inet6num} & 231,355 & 8,717 & (3.77\%) \\
 \texttt{organisation} & 82,512 & 0 & (0.00\%) \\
 \texttt{mntner} & 48,802 & 0 & (0.00\%) \\
 \texttt{aut-num} & 27,683 & 6,838 & (24.70\%) \\
 \texttt{role} & 20,684 & 14,430 & (69.76\%) \\
 \texttt{as-set} & 13,655 & 2,500 & (18.31\%) \\
 \texttt{route6} & 9,660 & 723 & (7.48\%) \\
 \texttt{irt} & 321 & 162 & (50.47\%) \\
 \midrule
 \textbf{Total} & \quad \textbf{5,239,201} & \quad \textbf{1,531,764} & \quad \textbf{(29.24\%)} \\
 \bottomrule
 \end{tabularx}
 \vspace{6pt}
 \caption{Data objects stored in the RIPE database, and references to DNS names.\newline 9 July, 2014.}
 \label{table:ripe_obj}
\end{table}
\setlength{\tabcolsep}{6pt}

Hijackable Internet resources are given by \texttt{inetnum} and \texttt{aut-num} objects, which represent blocks of IP addresses and unique numbers for autonomous systems respectively. Exemplary database objects are provided in \autoref{fig:ripe_entry}, further details on the RIPE database model and update procedures are available at \cite{ripespec}.

It is worth noting that the attacker neither needs authenticated access to the database nor does the attacker need to change the database objects. The attacker only needs to derive a valid contact point. We assume that the (publicly available) notification address usually belongs to the same DNS domain as the technical contact point. Detailed analysis is subject to future work; in our study, we disregard groups of objects that reference more than a single DNS domain as a precaution.

\subsubsection{Grouping Objects by Maintainer} \label{sec:grouping}

\begin{figure}[t!]
\centering
\includegraphics[width=0.74\textwidth]{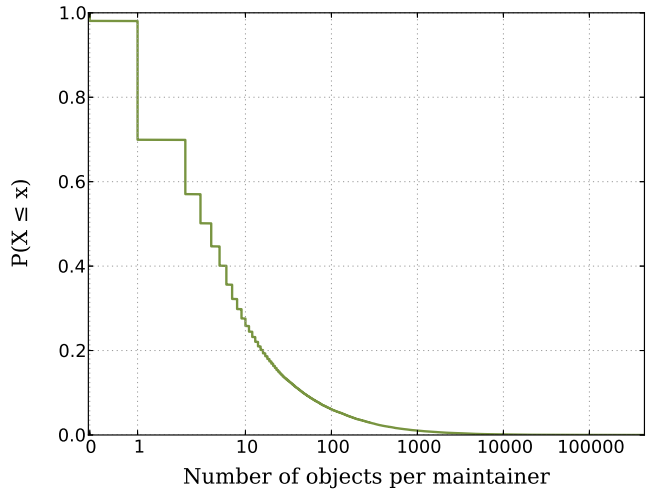}
\vspace{-9pt}
\caption{RIPE database objects grouped by references to a common maintainer object (CCDF).}
\label{fig:mntner_group_sizes}
\end{figure}

The RIPE database is mostly maintained by resource holders themselves. Its security model is based on references to \texttt{mntner} (maintainer) objects, which grant update and delete privileges to the person holding a \texttt{mntner} object's password. This security model allows us to infer objects under control of the same authority by grouping objects with references to a common \texttt{mntner} object. We use these \textit{maintainer groups} to estimate the impact of an attack for individual authorities: On average, we observed nearly 110 such references per \texttt{mntner} object, with a maximum of up to 436,558 references\footnote{The meta information refers to \textit{Interbusiness Network Administration Staff} of Telecom Italia.}. The distribution of the number of objects per maintainer group is presented in \autoref{fig:mntner_group_sizes}.

\begin{figure}[t!]
\begin{verbatim}
                inetnum:     194.28.196.0 - 194.28.199.255
                netname:     UA-VELES
                descr:       LLC "Unlimited Telecom"
                descr:       Kyiv
                notify:      internet@veles-isp.com.ua
                mnt-by:      VELES-MNT
\end{verbatim}
\begin{verbatim}
                aut-num:     AS51016
                as-name:     VALES
                descr:       LLC "Unlimited Telecom"
                notify:      internet@veles-isp.com.ua
                mnt-by:      VELES-MNT
\end{verbatim}
\vspace{-9pt}
\caption{Examples of RIPE database objects (\texttt{inetnum} and \texttt{aut-num} objects).}\label{fig:ripe_entry}
\end{figure}

For each of the maintainer groups, we obtain the set of all DNS names referenced by a group's objects. To unambiguously identify maintainer groups with expired domains, we merge disjoint groups that reference the same DNS domain, and discard groups with references to more than one DNS name. From an initial amount of 48,802 maintainer groups, we discard (a) 937 groups of zero size, i.e. unreferenced \texttt{mntner} objects, (b) 31,586 groups without domain name references, and (c) 4,990 groups with multiple references. The remaining 11,289 groups can be merged to 8,441 groups by identical DNS names. We further discard groups that do not include any hijackable resources, i.e. \texttt{inetnum} and \texttt{aut-num} objects, which finally leads us to 7,907 object groups. 

Note that the number of these groups is a lower bound: an attacker could identify even more with access to unanonymized RIPE data.
As discussed above, each of these groups is maintained by a single entity. If a group's DNS name expires, we consider the entity's resources to be a valuable target for an attacker. 

\subsection{Refinement by Activity Measures}
To confirm that a set of resources is abandoned, our approach is based on complementary data sources. We start with domain names that expire, which is a strong yet inconclusive indication for a fading resource holder. We gain further evidence by considering only resources that are neither changed in the RIPE database nor announced in BGP. Including both administrative (DNS, RIPE) and an operational (BGP) measures gives a comprehensive picture on the utilization of the resources.

\subsubsection{Lifetime of Domain Names}
\begin{figure}[!t] \centering
\includegraphics[width=0.74\textwidth]{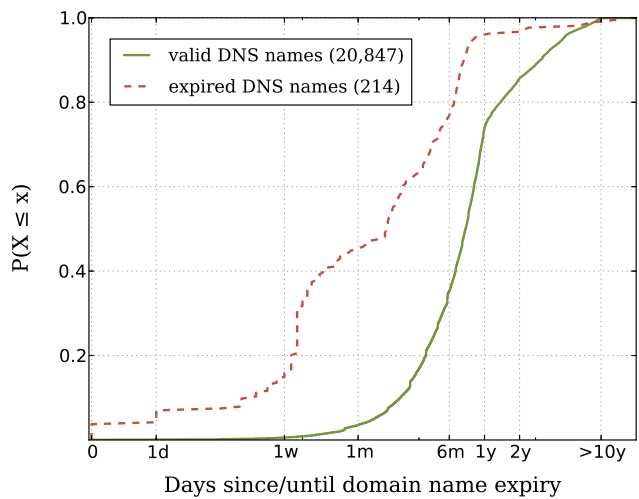}
\vspace{-10pt}
\caption{Expiry dates for DNS names referenced by RIPE database objects.}
\label{fig:domain_expiry_dist}
\end{figure}

We used the \textit{whois system} to query expiry dates for all extracted DNS names (cf., Section~\ref{sec:grouping}). \autoref{fig:domain_expiry_dist} shows the distribution of these dates. At the time of writing, 214 domain names have been expired. Another 121 names expire within the week, given that the owners miss to renew their contracts. The most frequent top level domains are \texttt{.com} (27.9\%), \texttt{.ru} (21.5\%), and \texttt{.net} (13.0\%), while the most frequent \textit{expired} TLDs are \texttt{.ru} (20.1\%), \texttt{.it} (16.4\%), and \texttt{.com} (9.81\%). The longest valid domains are registered until 2108 and mostly represent governmental institutions. The longest expired domain has been unregistered for nearly 14 years. With respect to the maintainer groups derived above, a total of 65 groups that reference expired DNS names remain. These groups hold 773 \texttt{/24} networks and 54 ASes, and are subject to closer investigation.

\begin{figure}[!t] \centering
\includegraphics[width=0.75\textwidth]{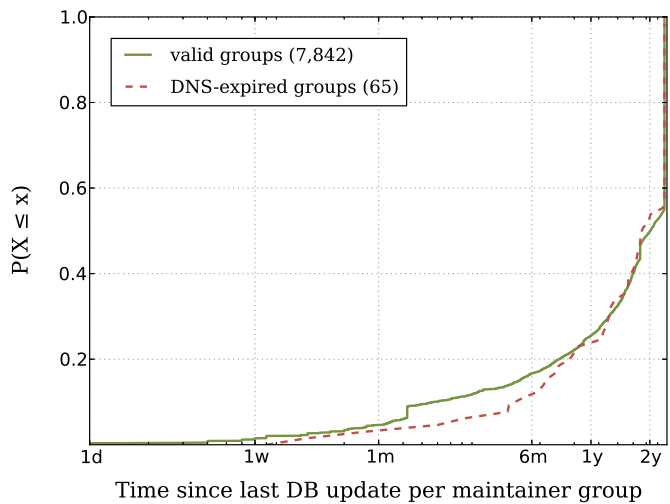}
\vspace{-10pt}
\caption{RIPE database updates by maintainer group (CDF).}
\label{fig:lastchange_ripedb}
\end{figure}

\subsubsection{RIPE Database Updates}
For each of the 7,907 maintainer groups -- divided into 7,842 valid groups and 65 with expired DNS names -- we extracted the minimum time since the last change for any of its database objects. Note that we filtered out automated bulk updates that affected \textit{all} objects of a certain type\footnote{For instance, RIPE added a new \texttt{status} attribute to all \texttt{aut-num} objects on 27 May, 2014.}. \autoref{fig:lastchange_ripedb} shows the distribution of database updates for groups with valid and for groups with expired domain names. While about 10\% of the valid groups show changes within two months, DNS-expired groups differ strikingly: the 10\%-quantile is at nearly 5~months. Hence, given these long times without updates, we consider resource groups that exhibit an object update within 6 months to be still maintained and not abandoned. Note that we do not assume inactivity in absence of such changes.

\subsubsection{BGP Activity}
\begin{figure}[!t] \centering
\includegraphics[width=0.75\textwidth]{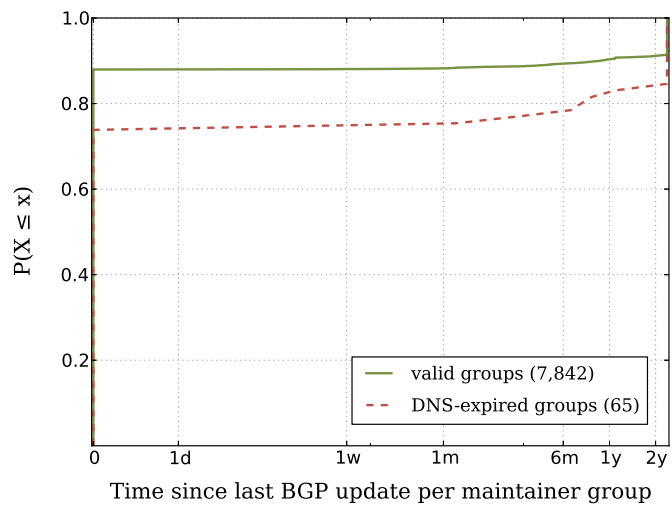}
\vspace{-10pt}
\caption{BGP update messages observed by maintainer group (CDF).}
\label{fig:lastchange_bgp}
\end{figure}

To confirm inactivity, we correlate the RIPE database updates with activities in the global routing system. For that, we analyze all available BGP update messages from the RouteViews Oregon's feed for the same time frame. This data set comprises 83,255 files with 18.4 billion announcements and 1.04 billion withdraw messages for resources assigned by RIPE. Given this data, we are able to extract two indicators: (1) the time since an \emph{IP prefix} was last visible from the RouteViews monitor, and (2) the time since the last deployment of a RIPE-registered \emph{AS number} by looking at AS path attributes. \autoref{fig:lastchange_bgp} shows the distribution of last activity in BGP for any Internet resource in our maintainer groups. Nearly 90\% of resources in valid groups are visible in BGP at the moment. Surprisingly, most of the remaining groups did not show any activity at all during the last 2.5 years. About 75\% of the DNS-expired resources are present in today's routing table -- and are thus still actively used. The remaining resources did show some activity in the past (10\%) or were never observed in BGP during our analysis period (15\%).

These findings confirm our assumption that inactivity in the RIPE database does not necessarily imply operational shutdown. While up to 85\% of the expired resources were seen in BGP within the last 2.5 years, \autoref{fig:lastchange_ripedb} indicates that not more than 55\% of the expired resources received an update in the RIPE database. We further learn that some expired resources did show BGP activity in the past, and do not show any activity today.
Note that we disregard resources with recent BGP activity. These resources could potentially be hijacked already; however, attacks that started before our analysis are beyond the scope of our approach.

\begin{figure}[!t] \centering
\includegraphics[width=0.75\textwidth]{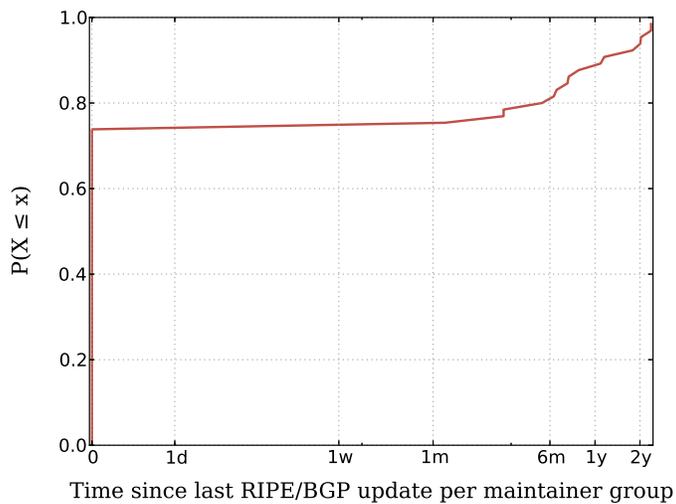}
\vspace{-10pt}
\caption{Combined RIPE/BGP activity by maintainer group (CDF).}
\label{fig:lastchange_all}
\end{figure}

\subsection{Hijackable Resources}
So far, we learned that 65 maintainer groups with a total of 773 \texttt{/24} networks and 54 ASes reference expired DNS names. Our activity measures further indicate that valid groups yield higher activity than expired groups. By combining these measures, we are able to infer resources that are inactive from both an administrative and an operational point of view. \autoref{fig:lastchange_all} shows the time since the latest change by any of these measures, i.e., the minimum value of both measures.

This combined activity measure clearly splits the 65 expired maintainer groups into two disjoint sets: 52 cases were active within the last 3 months, while 13~cases did not show any activity for more than 6 months. We consider these remaining 13 cases to be effectively abandoned. These resource groups represent a total number of 15 \texttt{inetnum} objects (with an equivalent of 73 \texttt{/24} networks) and 7 \texttt{aut-num} (i.e., AS number) objects. 

Now that we have identified vulnerable resources, we feel obliged to protect these resources. Since any attacker could repeat our analysis, we are going to contact endangered resource holders before publishing our findings. Although communication via e-mail is futile due to expired domains, we can fall back on telephone numbers provided in the RIPE database to reach out for the operators.

\section{Research Agenda} \label{sec:research} For the problem of abandoned
Internet resources, one might argue that the threat is not caused by a
technical but a social problem because operators agree to their peering
relations based on a weak authentication scheme. This scheme can be replaced by
stronger verification -- the required data already exists. RIRs have contracts
with the owners of delegated resources and thus are aware of more reliable
contact points (e.g., telephone numbers). However, the current situation shows
that we need mechanisms, tools, and procedures which are not tedious for
operators but allow for easy resource verification. Our approach to identify
abandoned resources can be easily extended to continuously monitor resources of
all RIRs. This would allow us to warn network operators about potential risks.
Finding scalable approaches to implement early warning and prevention in
real-time, though, is an open research issue.

\subsection{Limitations of Related Work} Current research is particularly
focused on the detection of BGP hijacking attacks. Proposed mitigation
techniques look on the control plane, the data plane, or both. \emph{Control
plane} monitoring is used to identify anomalies in BGP routing tables to infer
attacks \cite{phas,study,bogus,waehlisch,jacquemart14}. Such approaches are
prone to false positives due to legitimate causes for anomalies. Techniques
based on \emph{data plane} measurements account for changes of the router
topology \cite{hopcount,ispy}, or of hosts in supposedly hijacked networks
\cite{fingerprint,idle,argus}. These approaches rely on measurements carried
out before and during an attack. Beyond that, studies on the malicious intent
behind hijacking attacks exist \cite{flyby,spamtracer,ashijack,malicious}.

All detection approaches require the observation of suspicious routing changes.
Attacks based on our attacker model take place outside the routing system, and
thus do not lead to noticeable routing changes -- apart from a supposedly
legitimized organisation starting to reuse its Internet resources. Hence,
current detection systems are incapable to deal with this kind of attack.

The DNS has been widely studied in the context of malicious network activities,
mainly concerning spammers or fraud websites. Proactive blacklisting of domain
names \cite{fkp-ppdb-10} does not help in our scenario as the threat is
effective on the routing layer. Identifying orphaned DNS servers
\cite{kgccm-esods-10} is also out of scope of this paper as the attacker does
not leverage the DNS server but the expiring domain.

\subsection{Resource Ownership Validation} Despite its effectiveness, we
consider our approach to detect and monitor abandoned resources as outlined
above an intermediate solution only. In fact, we argue that there is a need for
resource ownership validation.

There is ongoing effort to increase the deployment of a \textit{Resource Public
Key Infrastructure (RPKI)} \cite{RFC-6480}. In its present state, the RPKI
allows for validation of route origins by using cryptographically secured
bindings between AS numbers and IP prefixes. This mechanism prevents common
hijacking attacks. In terms of hijacking abandoned resources, however, this
system is ineffective in its current form since the abandoned origin AS is
taken over as well, and origin validation performed by BGP routers
\cite{rfc-6811} will indicate a valid BGP update.

Even though the RPKI itself can be misused \cite{chbrg-rmra-13}, at the moment
it represents the only mechanism for proofing securely ownership of Internet
resources. We merely lack a clear procedure in the context of abandoned
Internet resources. One approach could be the following operational rule: a
peering request is only established when resource objects of the requesting
peer exist in the RPKI. Recent time stamps for these objects indicate that the
requesting peer has control over the resources as only authorized users can
create such objects.  Such a scheme seems feasible from the operational
perspective and might even increase the incentives to deploy RPKI.

RPKI is part of \textit{BGPsec}, an even larger effort to secure BGP. This
extension to the protocol remedies the risk of hijacking abandoned resources
due to its path validation capabilities: in our model, an attacker cannot
provide valid cryptographic keys to sign update messages as specified by
BGPsec~\cite{draft-bgpsec-spec}. However, the development of BGPsec is at an
early stage, and the benefit compared to pure origin validation is questionable
in particular in sparse deployment scenarios \cite{lgs-bspdj-13}.

Future research should be carried out on enabling Internet service providers to
validate resource ownership of customers. We see the potential of such a system
not only in preventing attackers from hijacking abandoned Internet resources.
It would also establish trust in customer-provider and peer-to-peer
relationships, as well as in resource transfers issued by RIRs or LIRs. 

\section{Conclusion} \label{sec:conclusion}

Motivated by a real-world case study, we introduced a generalized attacker
model that is aimed on the hijacking of abandoned Internet resources. We showed
that such an attack is feasible with little effort, and effectively hides the
attacker's identity by acting on behalf of a victim. By studying orthogonal
data sources over a period of more than 30 months, we could give evidence of a
high risk potential of such attacks. Only in the European RIR database, we
found 214 expired domain names that control a total of 773 \texttt{/24}
networks and 54 ASes, all of which can be easily hijacked. About 90\% of these
resources are still in use, which enables an attacker to disrupt operational
networks. The remaining 10\% of the resources are fully abandoned,
and ready to be stealthily abused.

Our findings led us to the conclusion that state-of-the-art systems are limited
to deal with this kind of attack. More importantly, we argued that there is a
need for \textit{resource origin validation}. Such a framework would not only
prevent attacks, but could also strengthen today's Internet eco-system by
establishing trust in resource ownership.

\clearpage
\subsubsection{Ethical Considerations} In this paper, we sketched a new attack
vector. Up until now, it is unclear how common such attacks are; our findings
thus might trigger new malicious activities. However, we also showed that this
attack is already known to attackers, and we sketched countermeasures to
mitigate this concern. In addition, we contact the holders of vulnerable
resources before publication of our findings.

\subsubsection{Acknowledgements} This work has been supported by the German
Federal Ministry of Education and Research (BMBF) under support code 01BY1203C,
project \textit{Peeroskop}, and by the European Commission under the FP7
project \textit{EINS}, grant number 288021.


\begin{thebibliography}{10}

\bibitem{study}
H.~Ballani, P.~Francis, and X.~Zhang.
\newblock A study of prefix hijacking and interception in the {I}nternet.
\newblock In {\em Proc. ACM SIGCOMM 2007}, pages 265--276, 2007.

\bibitem{chbrg-rmra-13}
D.~Cooper, E.~Heilman, K.~Brogle, L.~Reyzin, and S.~Goldberg.
\newblock {On the Risk of Misbehaving RPKI Authorities}.
\newblock In {\em Proc. of HotNets--XII}, New York, NY, USA, 2013. ACM.

\bibitem{fkp-ppdb-10}
M.~Felegyhazi, C.~Kreibich, and V.~Paxson.
\newblock {On the Potential of Proactive Domain Blacklisting}.
\newblock In {\em Proc. of the 3rd USENIX LEET Conference}, Berkeley, CA, USA,
  2010. USENIX Association.

\bibitem{idle}
{Hong, Seong-Cheol and Ju, Hong-Taek and Hong, James W.}
\newblock {IP Prefix Hijacking Detection Using Idle Scan}.
\newblock In {\em {12th Asia-Pacific Network Operations and Management
  Conference (APNOMS'09)}}, pages 395--404, 2009.

\bibitem{fingerprint}
X.~Hu and Z.~M. Mao.
\newblock Accurate real-time identification of {IP} prefix hijacking.
\newblock In {\em Proc. {IEEE} Symposium on Security and Privacy}, pages 3--17,
  2007.

\bibitem{jacquemart14}
Q.~{J}acquemart, G.~{U}rvoy {K}eller, and E.~{B}iersack.
\newblock {A} longitudinal study of {BGP} {MOAS} prefixes.
\newblock In {\em 6th {I}nt. {W}orkshop on {T}raffic {M}onitoring and
  {A}nalysis, (TMA '14)}, 2014.

\bibitem{kgccm-esods-10}
A.~J. Kalafut, M.~Gupta, C.~A. Cole, L.~Chen, and N.~E. Myers.
\newblock {An Empirical Study of Orphan DNS Servers in the Internet}.
\newblock In {\em Proc. of the 10th ACM SIGCOMM IMC}, pages 308--314, New York,
  NY, USA, 2010. ACM.

\bibitem{Kent2000}
S.~Kent, C.~Lynn, and K.~Seo.
\newblock {Secure Border Gateway Protocol (SBGP)}.
\newblock {\em IEEE Journal on Selected Areas in Communications}, 18(4), April
  2000.

\bibitem{phas}
M.~Lad, D.~Massey, D.~Pei, Y.~Wu, B.~Zhang, and L.~Zhang.
\newblock {PHAS: A prefix hijack alert system}.
\newblock In {\em Proc. 15th USENIX Security Symposium}, volume~15, 2006.

\bibitem{draft-bgpsec-spec}
M.~Lepinski.
\newblock {BGPSEC Protocol Specification}.
\newblock Internet-Draft -- work in progress~00, IETF, March 2011.

\bibitem{RFC-6480}
M.~Lepinski and S.~Kent.
\newblock {An Infrastructure to Support Secure Internet Routing}.
\newblock RFC 6480, IETF, February 2012.

\bibitem{lgs-bspdj-13}
R.~Lychev, S.~Goldberg, and M.~Schapira.
\newblock Bgp security in partial deployment: Is the juice worth the squeeze?
\newblock In {\em Proc. of ACM SIGCOMM}, pages 171--182, New York, NY, USA,
  2013. ACM.

\bibitem{rfc-6811}
P.~Mohapatra, J.~Scudder, D.~Ward, R.~Bush, and R.~Austein.
\newblock {BGP Prefix Origin Validation}.
\newblock RFC 6811, IETF, January 2013.

\bibitem{bogus}
J.~Qiu and L.~Gao.
\newblock {Detecting bogus BGP route information: going beyond prefix
  hijacking}.
\newblock In {\em In Proc. 3rd Int. Conf. on Security and Privacy in
  Communication Networks (SecureComm)}, 2007.

\bibitem{flyby}
A.~Ramachandran and N.~Feamster.
\newblock Understanding the network-level behavior of spammers.
\newblock In {\em Proc. ACM SIGCOMM 2006}, 2006.

\bibitem{ripespec}
{RIPE NCC}.
\newblock {RIPE Database Update Reference Manual}.
\newblock
  \url{http://www.ripe.net/data-tools/support/documentation/RIPEDatabaseUpdateManual20140425_edit.pdf}.

\bibitem{ashijack}
J.~Schlamp, G.~Carle, and E.~W. Biersack.
\newblock A forensic case study on {as} hijacking: the attacker's perspective.
\newblock {\em ACM SIGCOMM CCR}, 43(2):5--12, 2013.

\bibitem{argus}
X.~Shi, Y.~Xiang, Z.~Wang, X.~Yin, and J.~Wu.
\newblock Detecting prefix hijackings in the {I}nternet with argus.
\newblock In {\em Proc. ACM SIGCOMM Internet Measurement Conference (IMC)},
  2012.

\bibitem{spamtracer}
P.-A. {V}ervier and O.~{T}honnard.
\newblock {S}pam{T}racer: {H}ow stealthy are spammers?
\newblock In {\em 5th {I}nt. {W}orkshop on {T}raffic {M}onitoring and
  {A}nalysis, (TMA '13)}, 2013.

\bibitem{malicious}
{Vervier, Pierre-Antoine and Jacquemart, Quentin and Schlamp, Johann and
  Thonnard, Olivier and Carle, Georg and Urvoy-Keller, Guillaume and Biersack,
  Ernst W. and Dacier, Marc}.
\newblock {Malicious BGP Hijacks: Appearances can be deceiving}.
\newblock In {\em IEEE ICC Communications and Information Systems Security
  Symposium (ICC CISS 2014)}, 2014.

\bibitem{waehlisch}
M.~W\"{a}hlisch, O.~Maennel, and T.~C. Schmidt.
\newblock {Towards Detecting BGP Route Hijacking Using the RPKI}.
\newblock {\em ACM SIGCOMM CCR}, 42(4):103--104, August 2012.

\bibitem{ispy}
Z.~Zhang, Y.~Zhang, Y.~C. Hu, Z.~M. Mao, and R.~Bush.
\newblock {iSPY: Detecting IP prefix hijacking on my own}.
\newblock {\em IEEE/ACM Trans. on Networking}, 18(6):1815--1828, 2010.

\bibitem{hopcount}
C.~Zheng, L.~Ji, D.~Pei, J.~Wang, and P.~Francis.
\newblock A light-weight distributed scheme for detecting {IP} prefix hijacks
  in real-time.
\newblock In {\em Proc. ACM SIGCOMM 2007}, pages 277--288, 2007.

\end{thebibliography}
\end{document}